\documentstyle[twoside,fleqn,espcrc2]{article}

\def\sgline{\noalign{\vskip 0.10truecm\hrule\vskip 0.10truecm}}
\newcommand{\comment}[1]{}
\def\trhozm{T_\rho^{(2 {m})}}
\def\trhodm{T_\rho^{(3 {m})}}
\def\tkszm{T_{K^{\star}}^{(2 {m})}}
\def\tksdm{T_{K^{\star}}^{(3 {m})}}

\title{
{\begin{flushright}
       {\tt \small hep-ph/9612255\\
        TTP96-57\\
        HUTP-96/A058\\
        December 1996}
       \end{flushright}} 
 Theoretical Aspects of $\tau\rightarrow Kh(h)\nu_\tau$
  Decays\\ and Experimental Comparisons        }

\author{
M. Finkemeier \address{Lyman Laboratory of Physics,
        Harvard University,
        Cambridge, MA 02138, USA},
J.H.~K\"uhn and E.~Mirkes
\thanks{Talk presented by Erwin Mirkes at the Fourth International 
Workshop on Tau Lepton Physics (TAU 96), 16--19 September 1996, Estes Park,
Colorado, USA}
\address{Institut f\"ur Theoretische Teilchenphysik,
Universit\"at Karlsruhe, D-76128 Karlsruhe, Germany}}

\begin{document}
\begin{abstract}
Predictions for decay rates and distributions for
$\tau$ decays into final states with kaons are discussed
and compared with recent measurements.
Special emphasis is put on new constraints for the vector
current contribution in the $KK\pi$ decay modes.
For the $K\pi\pi$ modes, disagreements with the experimental 
results can be traced back to the $K_1$ widths.
\end{abstract}
\maketitle

\section{Introduction}
The $\tau$ lepton is heavy enough to decay into
a variety of hadronic final states. In particular, final states with kaons
can provide detailed information about low energy hadron physics 
in the strange sector. Topics to be studied include:\\
1) Chiral Perturbation Theory (CHPT) and effective Lagrangians,
2) Resonance parameters ($a_1,\rho,K_1,K^*,\ldots $) and radial excitations,
3) Tests of SU(3)$_F$ and isospin symmetry,
4) Structure of the charged hadronic current $[(V-A)$ or $(V+A)$],
   scalar contributions, etc.,
5) Determination of the strange quark mass and $\alpha_s(m_\tau)$
   measurements for $\Delta{S}=1$ transitions,
6) Measurement of the $\tau$ neutrino mass,
7) Search for CP violation effects beyond the Standard Model.

Predictions for final states with 2 and 3 meson final states involving
one or several kaons 
based on the ``chirally normalized vector meson dominance model''
will be discussed and compared with experimental
results.
The numbers for the experimental new world averages (NWA)
are taken from \cite{evans}.
Special emphasis is put 
on the vector part in the $KK\pi$ decay modes and 
on problems in the
axial vector part in the $K\pi\pi$ final states, which we believe can
be traced back to the $K_1$ widths.
The importance of a detailed analysis of the exclusive final
states with the structure function formalism is emphasized
\cite{km1}.

Let us first specify the general structure of the matrix elements for
semi-leptonic $\tau$ decays.
The  matrix element ${\cal{M}}$  for the hadronic $\tau$ decay 
into $n$ mesons $h_1, \ldots h_n$
\begin{equation}
\tau(l,s)\rightarrow\nu(l^{\prime},s^{\prime})
+h_{1}(q_{1},m_{1})+\ldots h_{n}(q_{n},m_{n}) \>,
\label{process2h}
\end{equation}
can be expressed in terms of a leptonic ($M_\mu$) and a
hadronic  current  ($J^\mu$) as
\begin{equation}
{\cal{M}}=\frac{G}{\sqrt{2}}\,
\bigl(^{\cos\theta_{c}}_{\sin\theta_{c}}\bigr)
\,M_{\mu}J^{\mu} \>.
\label{mdef2h}
\end{equation}
In Eq.~(\ref{mdef2h}), 
$G$ denotes the Fermi-coupling constant and   $\theta_c$ is the 
Cabibbo angle. The leptonic current is given by
\begin{equation}
M_{\mu}=
\bar{u}(l^{\prime},s^{\prime})\gamma_{\mu}(g_V-g_A\gamma_{5})u(l,s) \>,
\label{leptoncurrent}
\end{equation}
with $g_V=g_A=1$ in the Standard Model.
The hadronic current $J^{\mu}$ can  in general be  expressed in terms of
a vector and an axial vector current
\begin{eqnarray}
J^{\mu}(q_{1},\ldots,q_{n})&=&   \label{hadmat2h}\\
&&\hspace{-14mm}\langle h_{1}(q_{1})\ldots h_{n}(q_n)
|V^{\mu}(0)-A^{\mu}(0)|0\rangle \>.
\nonumber
\end{eqnarray}
In the following, we specify  the hadronic matrix elements
for hadronic decays into multi meson final states as expected from the
Standard Model.

\section{$\tau^-\rightarrow K^-\nu_\tau$}
The decay rate for the simplest decay mode with one kaon
is well
predicted by the kaon decay constants $f_K$
defined by the matrix element of the axial vector current
\begin{eqnarray}
\langle K(q) |A^{\mu}(0)|0\rangle \>   &=& i\sqrt{2}f_{K} q^{\mu}.
\end{eqnarray}
The kaon decay constant can be determined using the precisely measured
kaon decay widths $\Gamma(K\rightarrow \mu \bar{\nu}_\mu)$.
Radiative corrections $\delta R_{\tau/K}=(0.90\pm 0.22)\%$
to the ratios
$\Gamma(\tau\rightarrow K\nu)/\Gamma(K\rightarrow\mu\nu)$
have been calculated \cite{markus1}.
Using the recent world average $\tau_{\tau}=(291.6\pm1.6) fs$
for the tau lifetime   one obtains the following
theoretical predictions for the branching ratios
\begin{eqnarray}
{\cal{B}}(K\nu_\tau) &=& (0.723\pm 0.006)\% 
\end{eqnarray}
This prediction agrees well with the
new world average \cite{evans}
\begin{eqnarray}
{\cal B}^{exp}(K\nu_\tau) &=& (0.692\pm 0.028)\% 
\end{eqnarray}

\section{$\tau^-\rightarrow [K h ]^-\nu_\tau$}
We will now discuss the four decay modes
$\overline{K^0}\pi^-, K^-\pi^0, K^-\eta$ and $K^-K^0$.

\begin{figure}[tp]
\vspace{7.cm}
\begin{picture}(7,7)
\includegraphics{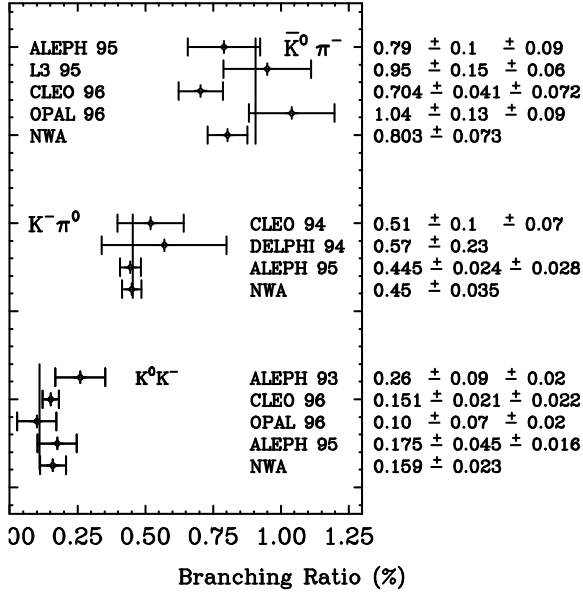}
\end{picture}
\vspace*{5mm}
\caption{
$\tau\rightarrow[Kh]^-\nu_\tau$ branching ratio measurements.
The vertical lines are the theoretical predictions from the second
column in table~\protect\ref{table1}.
}
\label{figbr_kpi}
\end{figure}

The hadronic matrix element for the decay
$\tau\rightarrow h_1 h_2 \nu_{\tau}$ 
can be 
expanded along a set of independent momenta
$q_1^{\mu}-q_2^{\mu}$ and $Q^\mu=(q_1^\mu+q_2^\mu)$
\begin{eqnarray}
\langle h_1(q_1) h_2(q_2) |V^{\mu}(0)|0\rangle \> &=&  \label{mat2h}\\
&&\hspace{-35mm}
[ (q_1-q_2)_\nu\,T^{\mu\nu}\,  F_V^{h_1 h_2} + Q^\mu\,F^{h_1 h_2}_S]
\nonumber
\end{eqnarray}
where $F_V (F_S)$ corresponds to the $J^P=1^-$ ($J^P=0^+$)
component of the weak charged current and
$T^{\mu\nu}$ is the transverse projector, defined by
\begin{equation}
T_{\mu\nu}=  g_{\mu \nu} - \frac{Q_\mu Q_\nu}{Q^2}  \>.
\label{trans}
\end{equation}
The $\overline{K^0}\pi^-, K^-\pi^0$ and $ K^-\eta$ decay modes are expected 
to be dominated by the $K^*$ resonance, whereas $K^-K^0$
is dominated by the $\rho$.
We allow for an admixture of radial excitations\footnote{The superscript $(2m)$
stands for ``2 meson'' resonances}:
\begin{equation} 
   \tkszm(Q^2)=\frac{ 
   \mbox{BW}_{K^\star}(Q^2) + \beta_{K^\star}\, \mbox{BW}_{{K^\star}'}(Q^2)
   }{1 + \beta_{K^\star}}
  \label{betakst}
\end{equation}
\begin{equation} 
   \trhozm(Q^2)  =  \frac{
   \mbox{BW}_\rho(Q^2) + \beta_\rho\, \mbox{BW}_{\rho'}(Q^2)
   }{1 + \beta_\rho}
   \label{beta}
\end{equation}
with \cite{fmkaon}
\begin{eqnarray}
&&\beta_{K^\star}=-0.135\>, \nonumber\\
&&m_{{K^\star}}=0.892 \, \mbox{GeV}\>,
\Gamma_{{K^\star}}=0.050 \mbox{GeV}\>,\\
&& m_{{K^\star}'}=1.412\, \mbox{GeV}\>, 
\Gamma_{{K^\star}'}=0.227\, \mbox{GeV}\>.\nonumber
\end{eqnarray}
and \cite{kuehn90}
\begin{eqnarray}
&&\beta_\rho = -0.145\>\nonumber\\
&&m_\rho = 0.773 \, \mbox{GeV}\>, \Gamma_\rho = 0.145 ,\mbox{GeV}\\
&&m_{\rho'} = 1.370 \, \mbox{GeV}\>, \Gamma_{\rho'} = 0.510 \, \mbox{GeV}\>.
\nonumber
\end{eqnarray}
In Eqs.~(\ref{betakst},\ref{beta}), BW denote
normalized Breit-Wigner propagators with an energy dependent 
width
\begin{equation}
\mbox{BW}_{X}[Q^2]\equiv {M^2_X\over [M^2_X-Q^2-i\sqrt Q^2 \Gamma_X(Q^2)]}\>.
\end{equation}
The vector form factors $F_V^{h_1h_2}$ in Eq.~(\ref{mat2h})
for the various 
two meson decay modes  are given by
\begin{eqnarray}
F_V^{\overline{K^0} \pi^-}    =
       \frac{1}{\sqrt{2}}\tkszm(Q^2) &&
F_V^{K^- \pi^0}     = \tkszm(Q^2)  \nonumber       \\
F_V^{K^- \eta}      =
        \sqrt{\frac{3}{2}} \tkszm(Q^2)  &&
F_V^{K^0 K^-}       = \trhozm(Q^2) \>.       \nonumber 
\end{eqnarray}
For the $\Delta S=1$ transition $\tau\rightarrow K\pi\nu_\tau$,
the form factor $F_S$ is expected to receive a sizable contribution 
($\sim 5\%$ to the decay rate) from the $K_0^*(1430)$
with $J^P=1^+$ \cite{fm95}.
However, we will neglect this contribution in the following discussion.
Predictions for the $K\pi$ and $KK$ decay modes are compared with experimental
results in Fig.~\ref{figbr_kpi}.
The sensitivity of our theoretical predictions
to the parameter $\beta_K^*$ and $\beta_\rho$
is indicated in table~\ref{table1}, which also includes a prediction for
the $K\eta$ final state. This decay mode is also completely fixed by the
parameters of the $\tkszm$ resonance.
\begin{table}[t]
\caption{
Braching ratios for two meson decay modes
}\label{table1}
\vspace{2mm}
\begin{tabular}{lcc}
        \hspace{1cm}
     & {$\beta_{K^*}=-0.135$ }
     & {$\beta_{K^*}= 0    $ }\\
\hline\\[-3mm]
$\overline{K^0}\pi^-$ & 0.906 \% & 0.65 \%   \\
$K^-\pi^0$       & 0.453 \% & 0.33 \%   \\
$K^-\eta$        & 2.0$\cdot10^{-2}$ \% 
                 & 0.8$\cdot 10^{-2}$ \% \\
                 \hline
        \hspace{1cm}
     & {$\beta_{\rho}=-0.145$ }
     & {$\beta_{\rho}= 0    $ }  \\
\hline \\[-3mm]
$K^-K^0 $ & 0.11 \% & 0.056 \%  \\[-5mm]
\end{tabular}
\end{table}
Note that our prediction for 
${\cal{B}}(K\eta)$ and
${\cal{B}}(KK)$ are both very sensitive to the choice of $\beta_{K^*}$
and $\beta_\rho$.
The results based on our favorite values  are
consitent with the experimental numbers.
The branching ratio  ${\cal B}(K\eta)$ was recently measured by CLEO
\cite{evans} ${\cal B}(K \eta) = (2.6\pm 0.5\pm 0.5)\cdot 10^{-2}\%$
and ALEPH \cite{evans}
${\cal B}(K \eta) = (2.9\pm 1.3\pm 0.7)\cdot 10^{-2}\%$.
Other theoretical predictions for the $K\eta$ decay mode
are ${\cal B}^{th}(K \eta) = (1.2-1.4)\cdot 10^{-2}\%$ \cite{pich_keta}
and ${\cal B}^{th}(K \eta) = 2.22\cdot 10^{-2}\%$ \cite{li_keta}.
Predictions for the $K^-K^0$ decay mode can also be obtained
via CVC from $e^+e^-\rightarrow\pi^+\pi^-$ data after applying
$SU(3)$-breaking effects. The result is 
${\cal B}^{CVC}(K^-K^0) = (0.111\pm 0.3)\%$ \cite{eidelman}.

\section{$\tau^-\rightarrow [K K \pi ]^-\nu_\tau$}
The hadronic matrix elements for three meson final states 
have a much richer structure. 
The decay modes involving  kaons allow
for axial and vector current contributions at the same time 
\cite{braaten,Dec93,fmkaon}.
The most general ansatz for the matrix element of the
quark current $J^{\mu}$  in Eq.~(\ref{hadmat2h})
is characterized by four form factors $F_i$ \cite{km1}, which 
are in general functions of
$s_1=(q_2+q_3)^2, s_2=(q_1+q_3)^2$, $s_3=(q_1+q_2)^2$
and $Q^2$ (which is conveniently chosen as an additional variable)
\begin{eqnarray}
J^{\mu}(q_{1},q_{2},q_{3})
&=&       \label{f1234}\\
&&  \hspace{-2cm}  V_{1}^{\mu}\,F_{1}
    + V_{2}^{\mu}\,F_{2}
    +\,i\, V_{3}^{\mu}\,F_{3}   
    + V_{4}^{\mu}\,F_{4} \>,
\nonumber
\end{eqnarray}
with
\begin{equation}
\begin{array}{ll}
V_{1}^{\mu}&= (q_{1}-q_{3})_{\nu}\,T^{\mu\nu}  \>,\\
V_{2}^{\mu}&= (q_{2}-q_{3})_{\nu}\,T^{\mu\nu}  \>,\\
V_{3}^{\mu}&= \epsilon^{\mu\alpha\beta\gamma}q_{1\,\alpha}q_{2\,\beta}
                                             q_{3\,\gamma} \>,
\\
V_{4}^{\mu}&=q_{1}^{\mu}+q_{2}^{\mu}+q_{3}^{\mu}\,=Q^{\mu} \>.
    \end{array}
\label{videf}
\end{equation}
$T^{\mu\nu}$ denotes again the transverse projector as defined 
 in Eq.~(\ref{trans}).
The form factors $F_{1}$ and $F_{2} (F_{3})$ originate from the 
$J^P=1^+$ axial vector
hadronic current ($J^P=1^-$ vector current) and correspond to a hadronic system
in a spin one state,
whereas $F_{4}$  is due to the $J^P=0^+$ spin zero part  of the 
axial current matrix
element.
The contribution to $F_4$ is expected to be  small
\cite{dfm} and we will neglect this contribution in the subsequent
discussion.

The form factors $F_1$  and $F_2$ can be predicted by chiral Lagrangians
at small momentum transfer whereas
the vector form factor is related 
to the Wess-Zumino anomaly \cite{WZ,kramer}.
To saturate the form factors in the region of large $Q^2$ by resonances with
momentum independent couplings is a natural choice in the
context of the vector dominance model (VDM)
but perhaps the most problematic assumption,
which has to be tested by measurements of differential distributions.
A particular powerful tool is provided by the analyses of
angular distributions. The relevant information is conveniently encoded in
structure functions \cite{km1}
which in turn allow to reconstruct the form factors.
$\tau$ decays are therefore a unique tool to study hadron physics in the
low momentum region in order to test a variety of theoretical approaches and
to derive resonance parameters which are not accessible elsewhere.

The resulting choice for the form factors $F_i$ 
in this chirally normalized vector meson dominance model 
is summarized by
\cite{fmkaon}
\begin{eqnarray}
F^{(abc)}_{1}(Q^2,s_2,s_3)\hspace{5mm}&=& \label{f1}\\
               &&\hspace{-2cm}\nonumber
                             {2\sqrt 2 A^{(abc)}\over 3f_\pi}
                              G_{1}^{(abc)}(Q^2,s_2,s_3) \>,
                              \\
F^{(abc)}_{2}(Q^2,s_1,s_3)\hspace{5mm}&=&\label{f2}\\
               &&\hspace{-2cm}\nonumber
                              {2\sqrt 2 A^{(abc)}\over 3f_\pi}
                              G_{2}^{(abc)}(Q^2,s_1,s_3) \>,
                              \\
F^{(abc)}_{3}(Q^2,s_1,s_2,s_3) &=& \label{f3}\\
               &&\hspace{-2cm}\nonumber
                              {A^{(abc)}\over 2\sqrt 2\pi^2f^3_\pi}  
                              G_3^{(abc)}(Q^2,s_1,s_2,s_3)\>.
\end{eqnarray}
where the Breit-Wigner functions $G_{1,2,3}$  and the normalizations
$A^{(abc)}$ are listed for $abc\equiv K\pi K$ in
Tab.~\ref{tabformkkpi}.

Let us briefly  discuss the resonance structure
in Tab.~\ref{tabformkkpi}
(for details see \cite{fmkaon}).

The form factors $F_1$ and $F_2$ are governed by the $a_1$
resonance  with energy dependent width
%
%
$$
  \mbox{BW}_{a_1}(Q^2) = 
  \frac{m_{a_1}^2}{m_{a_1}^2 - Q^2 - i m_{a_1} \Gamma_{a_1} g(Q^2) / 
  g(m_{a_1})}
$$
%
%
with
$
m_{a_1}       = 1.251 \,\, \mbox{GeV}\>, 
\Gamma_{a_1 } = \, 0.475\,\, \mbox{GeV}\>.
$
The  function $g(Q^2)$ has been calculated in
\cite{kuehn90}.

The two meson resonances $\trhozm$ and $\tkszm$ are defined in
Eqs.~(\ref{betakst}) and (\ref{beta}).
The $\omega$ resonance
$
T_\omega(Q^2) = \frac{1}{1 + \epsilon} [ \mbox{BW}_\omega(Q^2) +
\epsilon \mbox{BW}_\Phi(Q^2) ]
$
in the vector form factor $F_3$ in Tab.~\ref{tabformkkpi}
allows for a contribution of the $\phi$ with a relative strength
$\epsilon = 0.05\>$.

The admixture of the radial excitations
in the three meson non-strange vector resonance in $F_3$,
denoted by $\trhodm$, is expected to differ from the corresponding
two meson vector resonance $\trhozm$ in Eq.~(\ref{beta}).
In \cite{fmkaon,Dec93,tauola}, a form for $\trhodm$
including $\rho$, $\rho'$ and $\rho''$ was used, which was obtained from
a fit to (fairly poor) $e^+ e^- \to \eta \pi \pi$ data \cite{DM2,Gom90}.

\begin{figure}[tp]
\vspace{6.8cm}
\begin{picture}(7,7)
\includegraphics{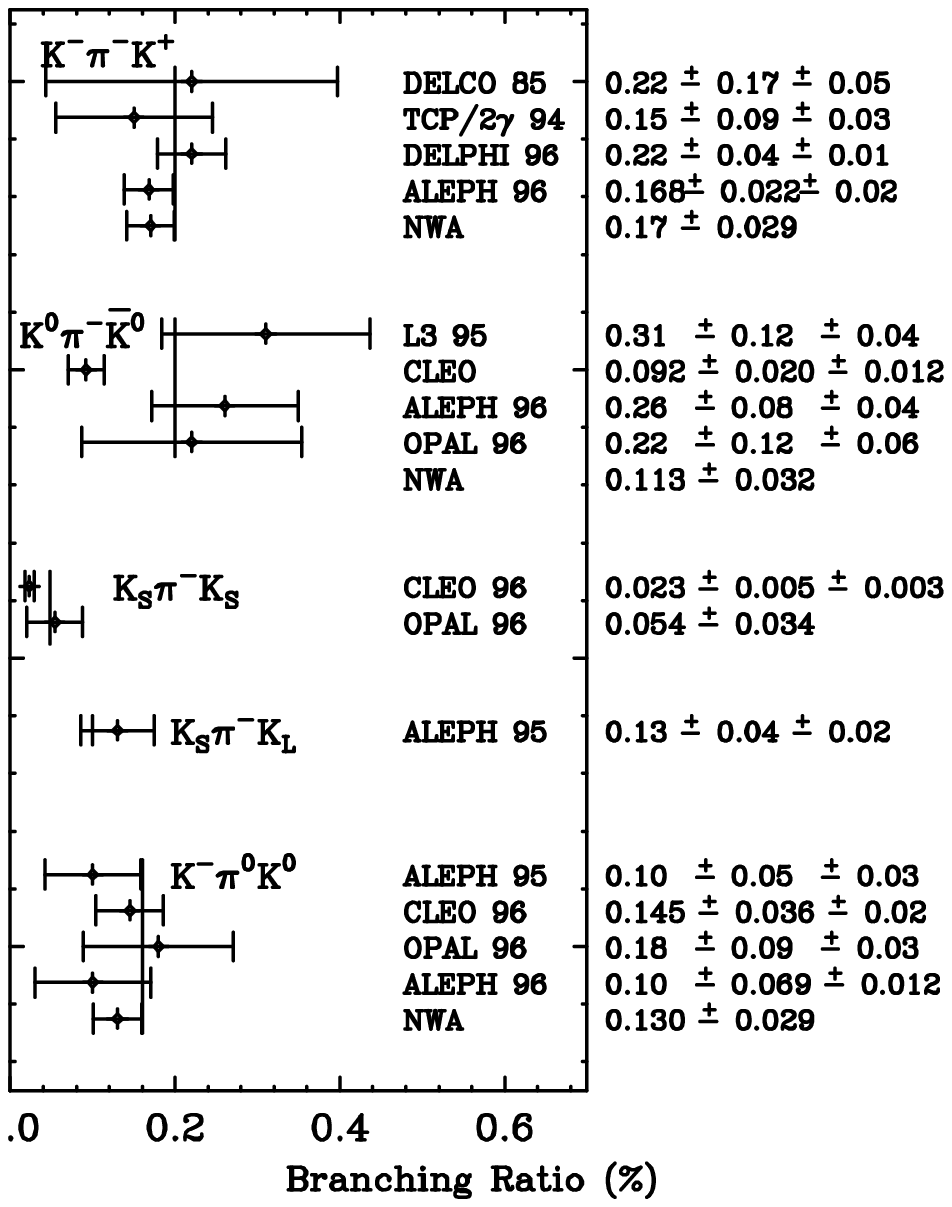}
\end{picture}
\vspace*{2.4cm}
\caption{
$\tau\rightarrow[K\pi K]^-\nu_\tau$ branching ratio measurements.
The vertical lines are the theoretical predictions from the second column 
in table~\protect\ref{tabkkpi}.
}
\label{figbr_kkpi}
\end{figure}

Predictions based on these parametrizations 
and the sub-resonance structure as given in 
table~\ref{tabformkkpi} are compared with
experimental results in Fig.~\ref{figbr_kkpi}.
The branching ratios are also listed in the second column of 
table~\ref{tabkkpi}.
\begin{table}
\caption{Predictions for the 
branching ratios ${\cal B}(abc)$ in $\%$ for the
$KK\pi$ decay modes. The contribution from the vector current
is given in parentheses.}
\label{tabkkpi}
$$
\begin{array}{c@{\quad}c@{\quad}c}
\hline \hline \\
\mbox{channel (abc)} &
\trhodm       &
\trhodm       \\
               &
\protect\cite{fmkaon}               &
\mbox{Eq.}(\protect\ref{trho_etapipi}).              
 \\ \hline
\\[-3mm]
K^- \pi^- K^+            & 0.20 \,(0.08)  &  0.18 \,(0.063)          \\ 
K^0 \pi^- \overline{K^0} & 0.20 \,(0.08 ) &  0.18 \,(0.063)          \\ 
K_S \pi^- K_S            & 0.048\,(0.014) &  0.045 \,(0.011)          \\ 
K_S \pi^- K_L            & 0.10 \,(0.052) &  0.092 \,(0.039)         \\
K^- \pi^0 K^0            & 0.16 \,(0.057) &  0.15  \,(0.045)         \\
\hline \hline
\end{array}
$$
\end{table}

\begin{figure}[tp]
\vspace{5.8cm}
\begin{picture}(7,7)
\includegraphics{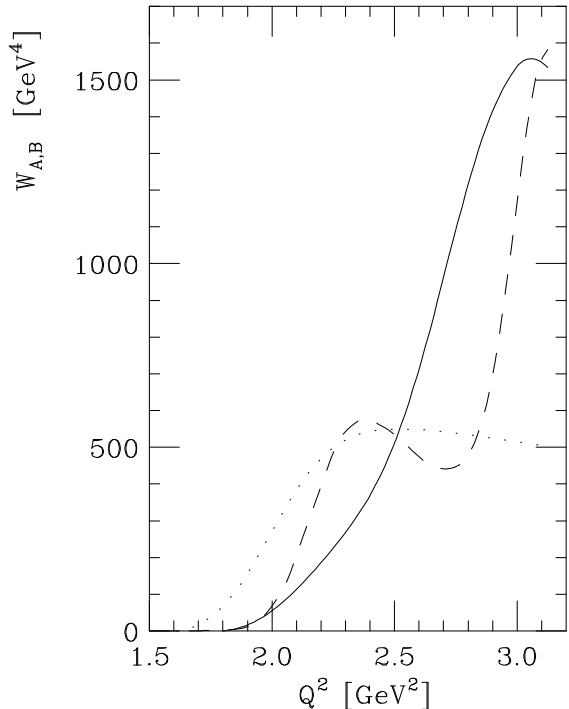}
\end{picture}
\vspace*{2.5cm}
\caption{
Axial vector structure function $w_A(Q^2)$ (dotted) and vector
structure function $w_B(Q^2)$ 
for the $\tau\rightarrow K^-\pi^- K^+\nu_\tau$ decay mode.
The solid line shows
$w_B$ based on the
$\trhodm$ parametrization in
Eqs.~(\protect{\ref{trho_etapipi}},\protect{\ref{bwetapipi}}),
whereas the dashed curve shows $w_B$ for
$\trhodm$ as defined in \protect\cite{fmkaon} 
(denoted as $T_\rho^{(2)}$ there).
}
\label{wab_kkpi}
\end{figure}

Using $SU(3)$ symmetry, $\trhodm$ can also be directly
obtained from $\tau\rightarrow \eta\pi^-\pi^0\nu_\tau$.
The matrix element for this decay mode is directly proportional
to the product $\trhodm\cdot \trhozm$ \cite{Dec93}.
New measurements of the $\eta\pi^-\pi^0$ mass spectrum in
$\tau\rightarrow \eta\pi^-\pi^0\nu_\tau$ have become available
\cite{cleo_omegapi,aleph_omegapi} allowing now for such a direct determination
of $\trhodm$.
We found, that the
$\eta\pi^-\pi^0$ mass spectrum 
in $\tau$ decays is poorly described by
$\trhodm$ parametrization based on the $e^+e^-\rightarrow\eta\pi\pi$ data
\cite{res}. 
A direct fit for the three body vector resonance (using the model
for $\tau\rightarrow\eta\pi^-\pi^0\nu_\tau$ in \cite{Dec93})
to the new $\tau\rightarrow \eta\pi^-\pi^0\nu_\tau$ 
data 
yields \cite{res}
\begin{equation}
\trhodm  =
\frac{  \Big[ \mbox{BW}_\rho + \lambda \mbox{BW}_{\rho'}
  + \nu \mbox{BW}_{\rho''}\Big] }{1 + \lambda + \nu}
\label{trho_etapipi}
\end{equation}
with
\begin{eqnarray}
\lambda &=& - 0.22 \pm 0.03  \nonumber \\
\nu &=& - 0.10 \pm 0.01\label{bwetapipi}\\
\chi^2/\rm{d.o.f.} &=& 11.0/14 \quad \nonumber
\end{eqnarray}
where we fix the parameters to the PDG \cite{RPP96} values,
\begin{eqnarray}
\begin{array}{ll}
m_\rho = 0.773 \mbox{ GeV}\>,    &\Gamma_\rho = 0.145 \mbox{ GeV}\>, \\ 
m_{\rho'} = 1.465 \mbox{ GeV}\>, &\Gamma_{\rho'} = 0.310 \mbox{ GeV}\>,\\ 
m_{\rho''} = 1.70 \mbox{ GeV}\>, &\Gamma_{\rho''} = 0.235 \mbox{ GeV}\>.
\end{array}
\label{masses}
\end{eqnarray}
The invariant mass distribution and the decay rate for the
$\tau\rightarrow \eta\pi^-\pi^0\nu_\tau$ are well described by these 
parameters \cite{res} and we will use this parametrization for
$\trhodm$ also in the following for the $KK\pi$ decay modes.
Predictions for the $KK\pi$ branching ratios based on these parameters
for $\trhodm$
are shown in the third column of table~\ref{tabkkpi}.
Note that the differences to the results in the second column are entirely
due to the different vector current contribution.
Whereas the differences in the branching ratios 
are fairly small
($\sim 10\%$) the differences in the $KK\pi$ mass spectra
are much larger.
Fig.~\ref{wab_kkpi} shows the $Q^2$ distribution for the
$s_1,s_2$ integrated structure functions $w_A(Q^2)$ and $w_B(Q^2)$
(for a definition of the structure functions see \cite{km1}).
Sizable differences are seen in the vector structure function
$w_B(Q^2)$ depending on the choice of the $\trhodm$
parametrization.

\begin{figure}[tp]
\vspace{5.8cm}
\begin{picture}(7,7)
\includegraphics{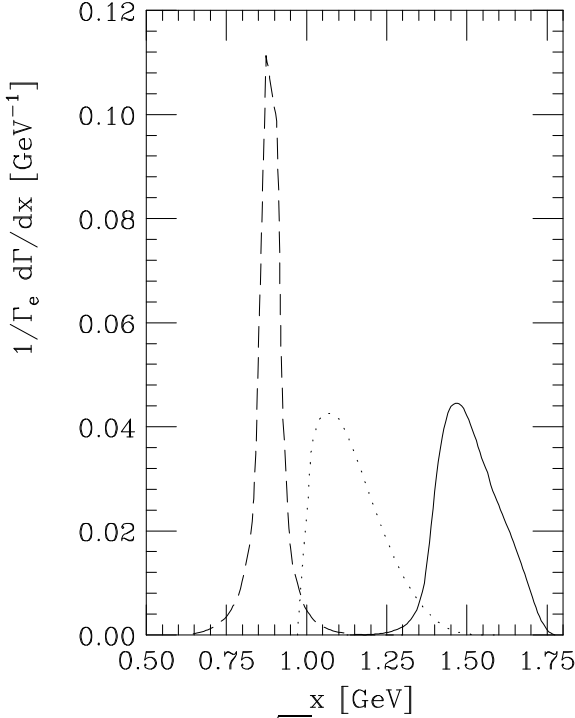}
\end{picture}
\vspace*{2.5cm}
\caption{
$x=\protect\sqrt{Q^2}=m(K^-\pi^-K^+) $ (solid), 
$x=\protect\sqrt{s_1}=m(K^+\pi^-)$ (dashed) and 
$x=\protect\sqrt{s_2}=m(K^-K^+)$ (dotted)
invariant mass  distributions
for the $\tau\rightarrow K^-\pi^- K^+\nu_\tau$ decay mode
normalized to $\Gamma_e$.
}
\label{wab_kkpi_all}
\end{figure}

The resonance structure in the $K^-\pi^-K^+$ decay mode
is shown in Fig.~\ref{wab_kkpi_all}.
The large peak in the $K^+\pi^-$
invariant mass around the $K^*(892)$ resonance shows  that these
decay modes are dominated by the $\tkszm$-two meson sub-resonance
compared to the  $\rho(\rightarrow K^+K^-)$
channel. The results are in good agreement with the measurements
in \cite{delphi}.

\begin{figure}[tp]
\vspace{5.8cm}
\begin{picture}(7,7)
\includegraphics{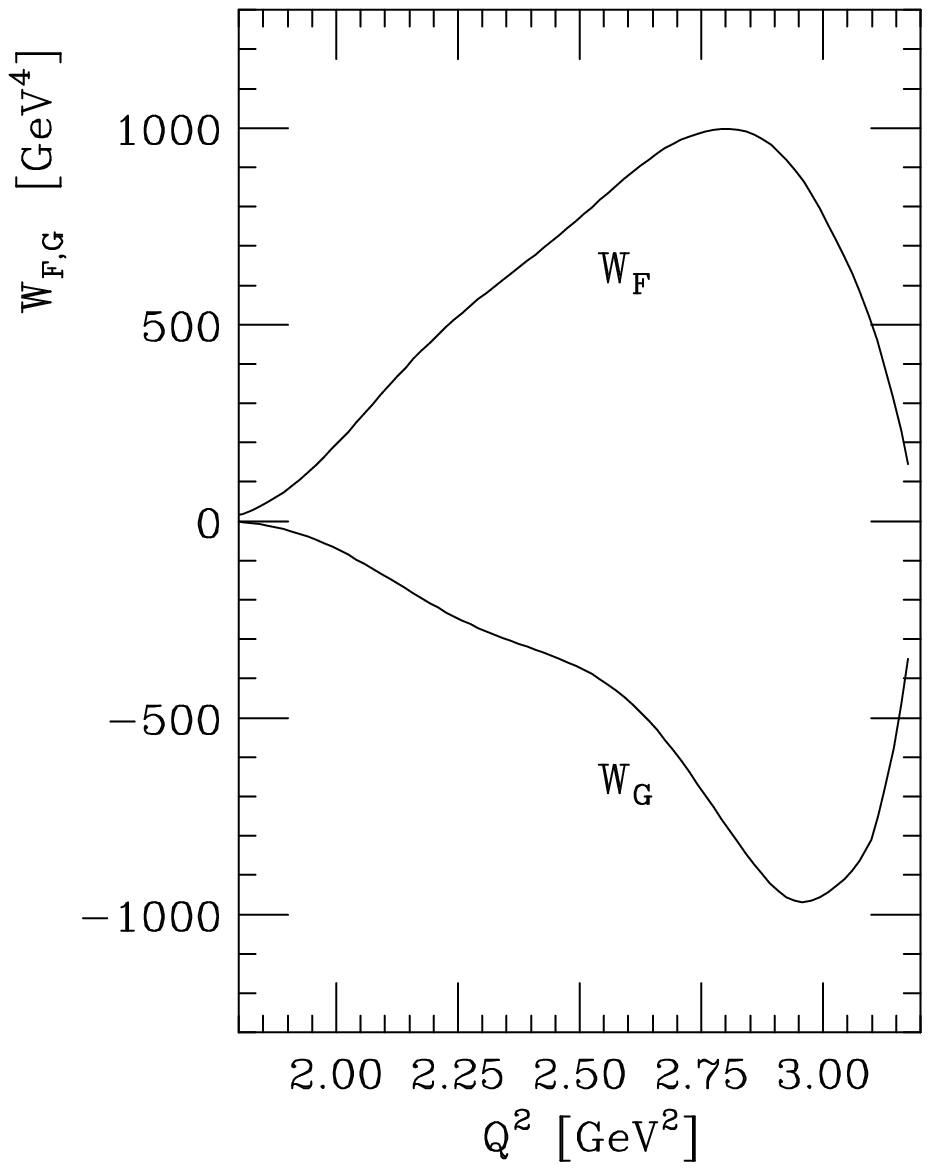}
\end{picture}
\vspace*{2.5cm}
\caption{
Axial vector-vector interference structure functions $w_F$ and
$w_G$ 
for the $\tau\rightarrow K^-\pi^- K^+\nu_\tau$ decay mode.
}
\label{wfg_kkpi}
\end{figure}

$\tau$ decay modes with an axial {\it and} a vector current
contribution offer a unique possibility to
measure the relative sign between $V$ and $A$ in the hadronic
current of Eq.~(\ref{hadmat2h}).
This is possible by just measuring the sign of one axial vector-vector
interference structure function $W_{F,G,H,I}$.
Predictions for the $Q^2$ distribution for the 
structure functions $w_{F,G}=\int ds_1 ds_2 W_{F,G}$ are
shown in Fig.~\ref{wfg_kkpi} for the
for the $\tau\rightarrow K^-\pi^- K^+\nu_\tau$ decay mode.
The size of these structure functions
is  comparable to the structure functions
$w_A$ and $w_B$ in Fig.~\ref{wab_kkpi} which determine the decay rate.
Further predictions for  $V$ and $A$ interference structure functions
in the $KK\pi$ and $K\pi\pi$ decay modes based on the model
in \cite{Dec93} are given in \cite{dm1}.


Structure function measurements allow also for a (model independent 
or model dependent) separation of vector, axial-vector and scalar contributions
in semi-leptonic $\tau$ decays \cite{ute}.
Furthermore, CHPT predicts interesting effects in the structure functions
$w_D$ and $w_E$ for the two three pion decay modes
\cite{long}.

Our result for 
${\cal{B}}(K_S\pi^-K_S)$ appears to be considerably higher than
the experimental result, whereas the other predictions agree fairly well.
The rates for $K_L\pi^-K_L$ and $K_S\pi^-K_S$ are identical,
the rate for $K_S \pi^- K_L$ is about a factor 2.1-2.4 higher.
Note that the first relation is a strict consequence of
CP symmetry, the second one depends on the dynamics of the decay
amplitude (in particular on the $a_1$ parameters \cite{fmkaon}).
Our results for the $K^-\pi^-K^+$ final state 
differ from those in \cite{Gom90}, where the contribution 
of the axial-vector channel amounts to less than 10\% 
to the decay rate in this channel.
In fact, our predictions for the axial-vector contribution is
about 60-75\% (see table~\ref{tabkkpi}).
This result is fairly insensitive towards the details of
the $\tkszm$ parametrization.
It is, however, sensitive towards the $a_1$ parameters.
Use of  $\Gamma_{a_1}=0.599$ GeV 
reduces the axial-vector contribution by about 15\%.

\begin{table*}[htbp]
  \setlength{\tabcolsep}{1.5pc}
  \caption{
Parametrization of the form factors
$F_1$ $F_2$ and $F_3$ in
 Eqs.~(\protect\ref{f1},\protect\ref{f2},\protect\ref{f3})
for $KK\pi$ decay modes.
}
\label{tabformkkpi} 
  \begin{tabular*}{\textwidth}{lccc}
  \sgline\sgline
$\begin{array}{c}
channel \\(abc)
\end{array}$ &
$A^{(abc)}$   & 
$G_1^{(abc)}(Q^2,s_2,s_3)$ &
$G_2^{(abc)}(Q^2,s_1,s_3)$                         \\
\sgline
$K^- \pi^- K^+$         & 
$\frac{- \cos \theta_c}{2}$ &
$\mbox{BW}_{a_1}(Q^2) \trhozm(s_2)$ &
$\mbox{BW}_{a_1}(Q^2) \tkszm (s_1)$   \\[2mm]
$K^0 \pi^- \overline{K^0}$ & 
$\frac{- \cos \theta_c}{2}$ &
$\mbox{BW}_{a_1}(Q^2) \trhozm(s_2)$ &
$\mbox{BW}_{a_1}(Q^2) \tkszm (s_1)$
\\[2mm]
$K_S \pi^- K_S$ &
$\frac{- \cos \theta_c}{4}$ &
$\mbox{BW}_{a_1}(Q^2) \tkszm(s_3)$ &
$\begin{array}{c}
\mbox{BW}_{a_1}(Q^2)\times \\{} 
[ \tkszm (s_1)
+ \tkszm(s_3) ]
\end{array}$
\\[4mm]
$K_S \pi^- K_L$ &
$\frac{- \cos \theta_c}{4}$ &
$\begin{array}{c}
\mbox{BW}_{a_1}(Q^2)
\times \\{} 
   [ 2 \trhozm(s_2) + \tkszm(s_3)]
\end{array}$
 &
$\begin{array}{c}
\mbox{BW}_{a_1}(Q^2)\times \\{} 
[ \tkszm (s_1)
- \tkszm(s_3) ]
\end{array}$
\\[4mm]
$K^- \pi^0 K^0$ &
$\frac{3 \cos \theta_c}{2 \sqrt{2}} $&
$\begin{array}{c}
\mbox{BW}_{a_1} (Q^2) \times \\{} 
\left[ \frac{2}{3} \trhozm(s_2)
+ \frac{1}{3} \tkszm(s_3) \right]
\end{array}$
 &
$\begin{array}{c}
\frac{1}{3}\mbox{BW}_{a_1} (Q^2) \times \\{} 
\left[  \tkszm(s_1)
-  \tkszm(s_3) \right] 
\end{array}$\\
  \end{tabular*}
  \begin{tabular*}{\textwidth}{lcc}
  \sgline
&
& 
$G_3^{(abc)}(Q^2,s_1,s_2,s_3)$
\\
\sgline
$K^- \pi^- K^+ $& 
$- \cos \theta_c$ &
$\trhodm(Q^2) (\sqrt{2} - 1) \left[ \sqrt{2} T_\omega(s_2)
+ \tkszm(s_1) \right]$
\\[2mm]
$K^0 \pi^- \overline{K^0}$ & 
$\cos \theta_c$ &
$\trhodm(Q^2) (\sqrt{2} - 1) \left[ \sqrt{2} T_\omega(s_2)
+ \tkszm(s_1) \right]$
\\[2mm]
$K_S \pi^- K_S$ &
$\frac{- \cos \theta_c}{2}$ &
$\trhodm(Q^2) (\sqrt{2} - 1) \left[  \tkszm(s_1)
- \tkszm(s_3) \right]$
\\[2mm]
$K_S \pi^- K_L $&
$\frac{ \cos \theta_c}{2}$ &
$\trhodm(Q^2) (\sqrt{2} - 1) \left[ 2 \sqrt{2} T_\omega(s_2)
+ \tkszm(s_1)
+ \tkszm(s_3) \right]$
\\[2mm]
$K^- \pi^0 K^0 $&
$\frac{- \cos \theta_c}{\sqrt{2}}$ &
$\trhodm(Q^2) (\sqrt{2} - 1) \left[  \tkszm(s_3)
- \tkszm(s_1) \right]$\\
\sgline
  \end{tabular*}
\end{table*}
\begin{table*}[htbp]
  \setlength{\tabcolsep}{1.5pc}
  \caption{
Parametrization of the form factors
$F_1$ $F_2$ and $F_3$ in
 Eqs.~(\protect\ref{f1},\protect\ref{f2},\protect\ref{f3})
for $K\pi\pi$ decay modes.
}
\label{tabformkpipi} 
  \begin{tabular*}{\textwidth}{lccc}
  \sgline\sgline
$\begin{array}{c}
channel \\(abc)
\end{array}$ &
$A^{(abc)}$   & 
$G_1^{(abc)}(Q^2,s_2,s_3)$ &
$G_2^{(abc)}(Q^2,s_1,s_3)$                         \\
\sgline
$\pi^0 \pi^0 K^-$ &
$\frac{\sin \theta_c}{4}$ &
$T_{K_1}^{(a)} (Q^2) \tkszm(s_2)$ &
$T_{K_1}^{(a)} (Q^2) \tkszm(s_1) $
\\[2mm]
$K^- \pi^- \pi^+$ &
$\frac{- \sin \theta_c}{2}$ &
$T_{K_1}^{(a)} (Q^2) \tkszm(s_2)$ &
$T_{K_1}^{(b)} (Q^2) T_{\rho}^{(1)}(s_1)$ 
\\[2mm]
$\pi^- \overline{K^0} \pi^ 0$ &
$\frac{3 \sin \theta_c}{2 \sqrt{2}}$ &
$\begin{array}{l}
\,\,\,\,\frac{2}{3} T_{K_1}^{(b)} (Q^2) \trhozm (s_2) \\[1ex]
+ \frac{1}{3} T_{K_1}^{(a)} (Q^2) \tkszm(s_3) 
\end{array}$ &
$\frac{1}{3} T_{K_1}^{(a)} (Q^2) \left[ \tkszm(s_1)
 - \tkszm(s_3) \right]$\\
  \end{tabular*}
  \begin{tabular*}{\textwidth}{lcc}
  \sgline
&
& 
$G_3^{(abc)}(Q^2,s_1,s_2,s_3)$
\\
\sgline
$\pi^0 \pi^0 K^-$ &
$\sin \theta_c$ &
$\frac{1}{4} \tksdm(Q^2) \left[ \tkszm(s_1) 
- \tkszm(s_2) \right]$
\\[2mm]
$K^- \pi^- \pi^+$ &
$ \sin \theta_c$ &
$\frac{1}{2} \tksdm(Q^2) \left[ \trhozm(s_1) 
+  \tkszm(s_2) \right]$
\\[2mm]
$\pi^- \overline{K^0} \pi^ 0$ &
$ \sqrt{2} \sin \theta_c$ &
$\frac{1}{4} \tksdm(Q^2) \left[2 \trhozm(s_2) 
+ \tkszm(s_1)
+ \tkszm(s_3) \right]$
\\ 
\sgline
  \end{tabular*}
\end{table*}
%

\section{$\tau^-\rightarrow [K \pi \pi ]^-\nu_\tau$}
The parametrization for the form factors $F_1,F_2,F_3$ 
in Eqs.~(\ref{f1}-\ref{f3}) for the $K\pi\pi$ decay modes are listed in 
table~\ref{tabformkpipi}. 

The form factors $F_1$ and $F_2$ are governed by the
$J^P=1^+$ three particle resonances with strangeness 
\begin{eqnarray}
   T_{K_1}^{(a)}(Q^2) & = &\\
&&\hspace{-2cm} \frac{1}{1 + \xi}
   \Big[ \mbox{BW}_{K_1(1400)}(Q^2) + \xi \mbox{BW}_{K_1(1270)} (Q^2)
   \Big] \>,
\nonumber \\[2mm] 
   T_{K_1}^{(b)}(Q^2) & = & \mbox{BW}_{K_1(1270)}(Q^2) 
\end{eqnarray}
with $\xi = 0.33$  \cite{fmkaon}
and \cite{RPP96} (all numbers in GeV)
\begin{equation}
\begin{array}{ll}
  m_{K_1}(1400) = 1.402 , 
& \Gamma_{K_1}(1400)= 0.174 \,\>,\\
  m_{K_1}(1270) = 1.270 \ \>,
 & \Gamma_{K_1}(1270)= 0.090 \>.\\
\end{array}
\label{k1para}
\end{equation}

The  three meson  vector resonance in the $1^-$ configuration
in the form factor $F_3$, denoted by $\tksdm$,
include the higher radial excitations
${K^\star}'$ and ${K^\star}''$ 
\begin{equation}
   \tksdm  = \frac{
 \mbox{BW}_{K^\star} + \lambda \mbox{BW}_{{K^\star}'}
  + \mu \mbox{BW}_{{K^\star}''}
}{1 + \lambda + \mu}
\end{equation}
with
\begin{equation}
\lambda =  -0.25,\,\,\,\,\,  \mu =  - 0.038 
\label{lambda_ks}
\end{equation}
and \cite{RPP96} 
\begin{eqnarray}
m_{{K^\star}} = 0.892 \, \mbox{GeV}\>, & & 
\Gamma_{{K^\star}} = \, 0.050 \mbox{GeV}\>,
\nonumber \\
m_{{K^\star}'} = 1.412\, \mbox{GeV}\>, & & 
\Gamma_{{K^\star}'} = 0.227\, \mbox{GeV}\>,
\label{kspara} \\
m_{{K^\star}''} = 1.714 \, \mbox{GeV}\>, & & 
\Gamma_{{K^\star}''} = 0.323 \, \mbox{GeV}\>.
\nonumber 
\end{eqnarray}
The parameters $\lambda$ and $\mu$
in Eq.~(\ref{lambda_ks}) are those from \cite{fmkaon}.
Based on $SU(3)$ symmetry one should rather use
$\lambda=-0.22$ and $\mu=-0.1$ as used in
Eq.~(\ref{trho_etapipi}).
However, the numerical significance of these
details is fairly small, because of the small vector channel
contribution in the relevant decay modes (about 10 \% or less) \cite{fmkaon}. 
We will therefore not discuss this problem here.

\begin{figure}[tbp]
\vspace{6cm}
\begin{picture}(7,7)
\includegraphics{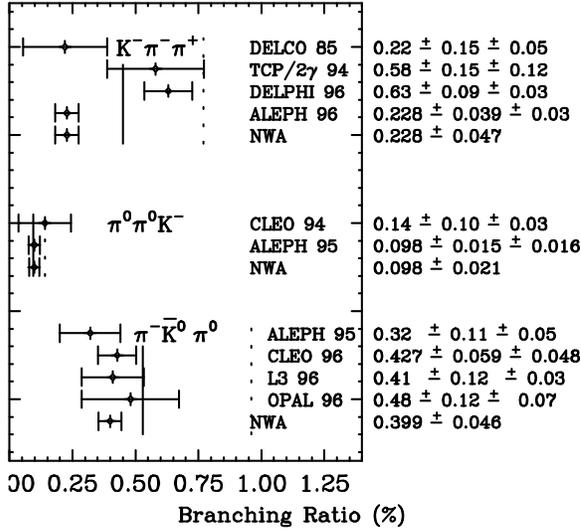}
\end{picture}
\caption{
$\tau\rightarrow[K\pi\pi]^-\nu_\tau$ branching ratio measurements.
The vertical dotted (solid)
lines are the theoretical predictions from the second (third)
column in table~\protect\ref{tab_kpipi}.
}
\label{fig_br_kpipi}
\end{figure}

Our predicitons for the branching ratios of the 
various $K \pi  \pi$ final states based on these parameters are
listed in the second column of table~\ref{tab_kpipi}.
The results 
are about a factor
of two larger than the experimental values
(see the vertical dotted lines in Fig.~\ref{fig_br_kpipi}.)

Moreover, QCD predicts \cite{qcd}
\begin{eqnarray}
   {\cal B}(\tau \to \nu_\tau + \mbox{hadrons}(J^P=1^+/0^-,\,S=-1)) &=&
\nonumber\\
\hspace{2cm}    (1.30 \pm 0.06)\%&&
\end{eqnarray}
for the inclusive decay rate into 
axial vector or pseudoscalar hadronic states with strangeness
$-1$. 
Subtracting from this the prediction for the branching ratio into a 
single kaon \cite{markus1}
\begin{equation}
   {\cal B}(\tau\to K \nu_\tau) = 0.72\%
\end{equation}
we find that the axial vector contribution to the three $K\pi\pi$ final
states must be less than $0.58 \%$. 
Our prediction  (second column in table~\ref{tab_kpipi}), 
however, for this contribution is
${\cal {B}}(\tau\to ( K\pi \pi)_A \nu_\tau) = 1.68 \%$.
So there is some strong indication that at least some of  our
predictions are by about a factor three too
large. 
\begin{table}
\caption{Predictions for the 
branching ratios ${\cal B}(abc)$ in $\%$ for the
$K\pi\pi$ decay modes.
Results are shown for $K_1$ paramaters in Eq.~(\protect\ref{k1para}) 
(second column, vector contribution in parentheses) and for 
$\Gamma_{K_1}(1400)= \Gamma_{K_1}(1270)=0.250 \,\,\mbox{GeV}$ 
(third column).}
\label{tab_kpipi}
$$
\begin{array}{c@{\quad}c@{\quad}c}
\hline \hline \\
\mbox{channel (abc)} &
      \Gamma_{K_1} \,\,\,[\mbox{Eq.}~(\protect\ref{k1para})]&
      \Gamma_{K_1}=0.250 \mbox{GeV}
 \\ \\
\hline
\\[-4mm]
\pi^0 \pi^0 K^-          & 0.14\,(0.012)  &    0.095             \\ 
K^-   \pi^-\pi^+         & 0.77\,(0.077)  &    0.45             \\ 
\pi^-\overline{K^0}\pi^0 & 0.96\,(0.010)  &    0.53             \\
\hline \hline
\end{array}
$$
\vspace*{-1.cm}
\end{table}

We believe that we have identified the widths of the $K_1$ particles as the
culprit. As mentioned before, 
the results in the second column in table~\ref{tab_kpipi} are
based on  the particle data group values for the 
widths of the two $K_1$ resonances  (see Eq.~(\ref{k1para})).
We believe that these numbers are considerably too small, maybe by factors 
of two or three.
We have three independent reasons to justify this statement.

1.) This is the only natural explanation we can find
      for the factor three discrepancy
      between the inclusive QCD constraint and our prediction in \cite{fmkaon}
      for the rate of $\tau \to (K\pi\pi)_A \nu_\tau$.
      Note that most of the other predictions of the
      chirally normalized vector meson dominance model
      agree fairly well with data.
%

2.)   From $SU(3)$ flavour symmetry and $\Gamma_{a_1} \approx 
      400 \cdots 600\, \mbox{MeV}$, the widths given in \cite{RPP96}
      seem unnaturally small.

3.) Until now, the widths of the $K_1$'s have only been measured in
      hadronic production. These measurements have strong backgrounds,
      and results for the resonance parameters depend on the 
      model used for the background. Remember that hadronic production
      of $a_1$ yielded small values for its widths, of about $300 \,
      \mbox{MeV}$, incompatible with results from $\tau$ decays.
      In fact, it has been shown in \cite{Bow88} that by a 
      modification of the coherent background 
      in the diffractive hadronic prodution of $a_1$, a considerably
      larger width can be extracted which is compatible with $\tau$
      data. 
      In the  $K_1$ measurements in \cite{Dau81b}, the same
      assumptions have been made as in \cite{Dau81}, which yielded
      those small values for the $a_1$ width.

\begin{figure}[tp]
\vspace{5.8cm}
\begin{picture}(7,7)
\includegraphics{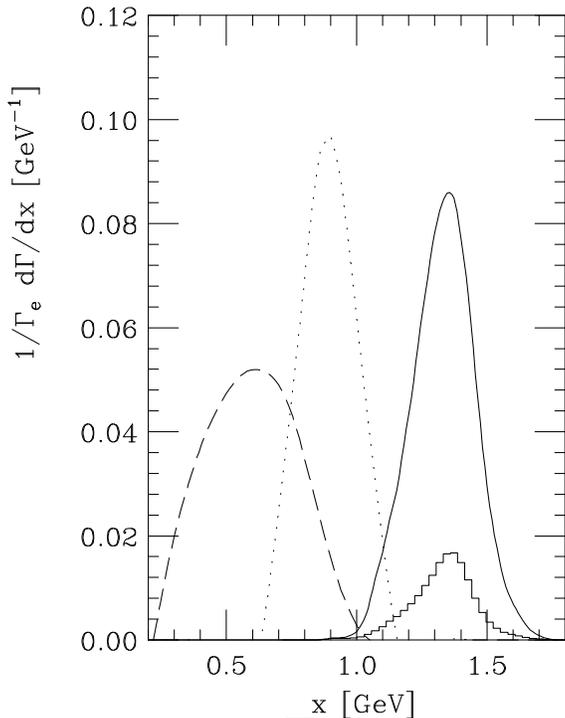}
\end{picture}
\vspace*{2.5cm}
\caption{
$x=\protect\sqrt{Q^2}=m(K^-\pi^-\pi^+)$ (solid), 
$x=\protect\sqrt{s_1}=m(\pi^+\pi^-)$ (dashed) and 
$x=\protect\sqrt{s_2}=m(K^-\pi^+)$ (dotted)
invariant mass  distributions
for the $\tau\rightarrow K^-\pi^- \pi^+\nu_\tau$ decay mode
normalized to $\Gamma_e$.
The results are based on 
$\Gamma_{K_1}(1400)= \Gamma_{K_1}(1270)=0.250 \, \mbox{GeV}$.
The contribtion from the vector part $(\sim |F_3|^2)$ to
the $K^-\pi^-\pi^+$ invariant mass distribution is shown
as the histogram.
}
\label{wab_kpipi}
\end{figure}
\begin{figure}[tp]
\vspace{5.8cm}
\begin{picture}(7,7)
\includegraphics{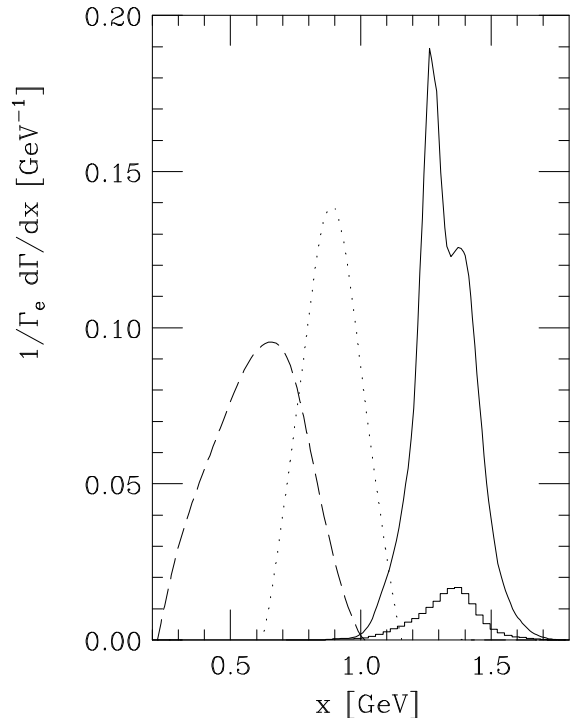}
\end{picture}
\vspace*{2.5cm}
\caption{
Same as Fig.~\protect\ref{wab_kpipi} but with the $K_1$
parameters from Eq.~(\protect\ref{k1para}).
}
\label{wab_kpipi_k1old}
\end{figure}
The strong sensitivity to the $K_1$ width  is demonstrated by the
numbers  in the third column of table~\ref{tab_kpipi}, where
predictions for the branching ratios 
based on $\Gamma_{K_1}=0.250$ GeV are shown.
The results are now much closer to the measured values
(see the vertical solid lines in Fig.~\ref{fig_br_kpipi}).
A direct measurement of the $K_1$ parameters in $\tau$ decays, in particular
a measurement of the widths, would be very interesting in view of
these results.

Finally, Figs.~\ref{wab_kpipi} (\ref{wab_kpipi_k1old})
show the
resonance structure in the $K^-\pi^-\pi^+$ decay mode
based on the two choices for the $K_1$ widths.
The results in Fig.~\ref{wab_kpipi} are in good agreement
with \cite{delphi}.
Fig.~4 in \cite{long} shows the structure
function $w_B$, which is very sensitive to  
the two choices of the $K_1$ widths.\\[2mm]
\section*{Acknowledgments}
We thank H. Evans for making his results for the latest
new world averages available prior to this conference.
We would also like to thank the conference organizers,
in particular Jim Smith, for the very well organized workshop.
The work of M.F.  was supported by the National Science
Foundation (Grant PHY-9218167) and by the Deutsche 
Forschungsgemeinschaft.
The work of E.~M. was supported in part  
by DFG Contract Ku 502/5-1.
%

%



\begin{thebibliography}{99}
\bibitem{evans}
H. Evans, in: Proceedings of the Fourth Workshop on Tau Lepton Physics
(TAU 96), to be published as Proceedings Supplement by Nucl. Phys. B

\bibitem{km1}
J.H. K\"uhn, E. Mirkes, Phys. Lett. B286 (1992) 381; Z. Phys. C56 (1992)
661; erratum {\it ibidem} C67 (1995) 364.

%


\bibitem{markus1} 
R.~Decker and M.~Finkemeier, 
Phys. Lett. B {\bf 334}, 199 (1994);
Nucl. Phys. B {\bf 438}, 17 (1995).


\bibitem{fmkaon}
M. Finkemeier, E. Mirkes, Z. Phys. C69 (1996) 243.

\bibitem{kuehn90}
J.H. K\"uhn, A. Santamaria, Z. Phys. C48 (1990) 445.


\bibitem{fm95}
M. Finkemeier, E. Mirkes, hep-ph/9601275, to be published in
Z. Phys. C.

%

\bibitem{pich_keta}
A. Pich, Phys. Lett. B {\bf 196}, 561 (1987).

\bibitem{li_keta}
B.A. Li, [hep-ph/9606402].

\bibitem{eidelman}
S.I. Eidelman, in: Proceedings of the Fourth Workshop on Tau Lepton Physics
(TAU 96), to be published as Proceedings Supplement by Nucl. Phys. B

\bibitem{braaten}
E. Braaten, R.J. Oakes, Int. J. Mod. Phys. A5 (1990) 2737.

\bibitem{Dec93}
R. Decker, E. Mirkes, R. Sauer, Z. Was, Z. Phys. C58 (1993) 445.


\bibitem{dfm}
R.~Decker, M.~Finkemeier and E.~Mirkes, Phys. Rev. D {\bf 50 }, 3197 (1994).


\bibitem{WZ} J.~Wess and B. Zumino, Phys. Lett B {\bf 37}, 95 (1971);\\
             E.~Witten, Nucl. Phys. B {\bf 223}, 422 (1983);
                          {\it ibid.} {\bf 223}, 433 (1983).


\bibitem{kramer}
G. Kramer, W.F. Palmer and S. Pinsky, Phys. Rev. D {\bf 30}, 89 (1984);\\
G. Kramer, W.F. Palmer, Z. Phys. C {\bf 25}, 195 (1984);
                    {\it ibid.}  C {\bf 39}, 423 (1988).

\bibitem{tauola}
S.~Jadach, J.H.~K\"uhn and Z.~Was,
Computer Phys. Comm. 64 (1991) 275;\\
R. Decker, J.H. K\"uhn, S. Jadach and  Z.~Was,
Computer Phys. Comm. 76 (1993) 361.

\bibitem{DM2}
DM2 Collaboration, A. Antonelli et al., Phys. Lett. B {\bf 212}, 133 (1988).

\bibitem{Gom90}
J.J. Gomez-Cadenas, M.C. Gonzales-Garcia and A. Pich, Phys. Rev. D {\bf 42},
3093 (1990).

\bibitem{cleo_omegapi}
CLEO Collaboration (R.~Balest {\it et al.}), Phys. Rev. Lett {\bf 75}
3809 ({1995}).

\bibitem{aleph_omegapi}
ALEPH Collaboration (D.~Buskulic {\it et al.}), {\it A study of $\tau$
decays involving $\eta$ and $\omega$ mesons}, CERN-PPE/96-103.

\bibitem{res}
E.~Mirkes and R.~Urech, Isospin Violation in $\tau\rightarrow 3\pi\nu_\tau$,
in preparation.

\bibitem{RPP96}
Review of Particle Physics, Particle Data Group,  Phys. Rev. D 54 (1996) 1.

\bibitem{delphi}
DELPHI coll., W. Hao. et al., DELPHI 96-76 Conf. 8, ICHEP'96 (pa07-009).

\bibitem{dm1}
R. Decker and E. Mirkes, 
Phys. Rev. D 47 4012 (1993).

\bibitem{ute}
U. M\"uller, in: Proceedings of the Fourth Workshop on Tau Lepton Physics
(TAU 96), to be published as Proceedings Supplement by Nucl. Phys. B;
and references therein.

\bibitem{long}
G. Colangelo, M. Finkemeier, E. Mirkes and R. Urech, 
hep-ph/9611329, in:
Proceedings of the Fourth Workshop on Tau Lepton Physics
(TAU 96), to be published as Proceedings Supplement by Nucl. Phys. B
for more details see:
G. Colangelo, M. Finkemeier,  and R. Urech, Phys. Rev. D54 (1996) 4403.


\bibitem{qcd}
E. Braaten, S. Narison, A. Pich, Nucl. Phys. B 373 (1992) 581;
S. Narison, A. Pich, Phys. Lett. B 304 (1993) 359;
M. Suzuki, Phys. Rev. D 49 (1994) 2634.

\bibitem{Bow88} 
M. G. Bowler, Phys. Lett. B 209 (1988) 99.

\bibitem{Dau81b}
ACCMOR Collaboration (C. Daum et al.), Nucl. Phys. B 187 (1981) 1.

\bibitem{Dau81}
ACCMOR Collaboration (C. Daum et al.), Nucl. Phys. B 182 (1981) 269.


\clearpage
\end{thebibliography}
\end{document}